\documentclass[article]{aa}

\usepackage{amsmath, amssymb, amsfonts, amscd}
\usepackage{mathtools} 
\usepackage{array} 
\usepackage[varg]{txfonts} 
\usepackage{color}
\usepackage{mathabx}
\usepackage{graphicx} 

\usepackage{natbib}
\bibpunct{(}{)}{;}{a}{}{,} 

\usepackage[colorlinks=true,linkcolor=blue,citecolor=blue]{hyperref}
\hypersetup{breaklinks=true}

\definecolor{emerald}{rgb}{0.3,0.85,0.2}
\newcommand{\mybf}[1]{#1}
\newcommand{\pcor}[1]{#1}
\newcommand{\cora}[1]{#1}

\pdfpageattr{/Group <</S /Transparency /I true /CS /DeviceRGB>>}

\begin{document}

\title{The rotation of planets hosting atmospheric tides: \\ from Venus to habitable super-earths}

\author{
P. Auclair-Desrotour\inst{1,2}
\and J. Laskar\inst{1}
\and S. Mathis\inst{2,3}
\and A. C. M. Correia\inst{1,4}
}

\institute{IMCCE, Observatoire de Paris, CNRS UMR 8028, PSL, 77 Avenue Denfert-Rochereau, 75014 Paris, France
\and Laboratoire AIM Paris-Saclay, CEA/DRF - CNRS - Universit\'e Paris Diderot, IRFU/SAp Centre de Saclay, F-91191 Gif-sur-Yvette Cedex, France
\and LESIA, Observatoire de Paris, PSL Research University, CNRS, Sorbonne Universit\'es, UPMC Univ. Paris 06, Univ. Paris Diderot, Sorbonne Paris Cit\'e, 5 place Jules Janssen, F-92195 Meudon, France
\and CIDMA, Departamento de F\'isica, Universidade de Aveiro, Campus de Santiago, P-3810-193 Aveiro, Portugal \\
\email{pierre.auclair-desrotour@obspm.fr; jacques.laskar@obspm.fr; stephane.mathis@cea.fr; correia@ua.pt} 
}

\date{Received ... / accepted ...}

\abstract{
{The competition between the torques induced by solid and thermal tides drives the rotational dynamics of Venus-like planets and super-Earths orbiting in the habitable zone of low-mass stars.}
The resulting torque determines the possible equilibrium states of the planet\mybf{'s} spin.
{We compute here an analytic expression  for the total tidal torque exerted on a Venus-like planet. This expression is used to characterize the equilibrium rotation of the body.    }
{Close to the star, the solid tide dominates. Far from it, the thermal tide drives the rotational dynamics of the planet. The transition regime corresponds to the habitable zone, where prograde and retrograde equilibrium states appear.}
{We demonstrate the strong impact of the atmospheric properties and of the rheology of the solid part on the rotational dynamics of Venus-like planets, highlighting the key role played by dissipative mechanisms in the stability of equilibrium configurations.}
}

\titlerunning{The rotation of planets hosting atmospheric tides}
\authorrunning{Auclair-Desrotour et al.}

\maketitle


\section{Introduction}

Twenty years after the discovery of the first exoplanet around a Solar-type star \citep{MQ1995}, the rapidly growing population of detected low-mass extra-solar planets has become considerable 
and their number will keep increasing, which generates a strong need \mybf{for} theoretical modeling and predictions. Because the mass of such planets is not large enough to allow them to accrete a voluminous gaseous envelope, they are mainly composed of a telluric core. In many cases \citep[e.g. 55 Cnc e, see][]{Demory2016}, this core is \mybf{covered} by a thin atmospheric layer as observed on the Earth, Venus and Mars. The \mybf{rotation} of these bodies strongly affects their surface temperature equilibrium and atmospheric global circulation \citep[][]{FL2014}. Therefore, it is a key quantity to understand their climate, particularly in the \mybf{so-called} ``habitable zone'' \mybf{\citep[][]{Kasting1993}}. 

The rotational dynamics of super-Venus and Venus-like planets is driven by the tidal torques exerted both on the rocky and atmospheric layers \citep[see][]{Correia2008}. The solid torque, which is induced by the gravitational tidal forcing of the host star, tends to despin the planet to pull it back to synchronization. The atmospheric torque is the sum of two contributions. The first one, caused by the gravitational tidal potential of the star\mybf{,} acts on the spin in a similar way as the solid tide. The second one, called ``thermal tide'', results from the perturbation due to the heating of the atmosphere by the stellar bolometric flux. The torque induced by this component is in opposition with those gravitationally generated. Therefore, it pushes the angular velocity of the planet away from synchronization. Although the mass of the atmosphere often represents a negligible fraction of the mass of the planet (denoting $ f_{\rm A} $ this fraction, we have $ f_{\rm A} \sim 10^{-4} $ for Venus), thermal tides can be of the same order of magnitude and even stronger than solid tides \citep{DI1980}. 

This competition naturally gives birth to prograde and retrograde rotation equilibria in the semi-major axis range defined by the habitable zone, in which Venus-like planets are submitted to gravitational and thermal tides of comparable intensities \citep{GS1969,DI1980,CL01}. Early studies of this effect were based 
on analytical models developed for the Earth \citep[e.g.][]{CL70} that present a singularity at synchronisation, while \citet{CL01} avoid this drawback by a smooth interpolation annuling the torque at synchronization.
Only recently, the atmospheric tidal perturbation has been computed numerically with Global Circulation Models (GCM) \citep[][]{Leconte2015}, and analytically \citep{ADLM2016} (P1), who generalized the reference work of \cite{CL70} by including the dissipative processes (radiation, thermal diffusion) that regularize the behaviour of the atmospheric tidal torque at the synchronization.

Here, we revisit the equilibrium rotation of super-Earth planets based on the atmospheric tides model presented in P1. For the solid torque, we use the simplest physical prescription, a Maxwell model \citep{RMZL2012,Correia2014}, because the rheology of these planets is unknown.

\section{Tidal torques}

\subsection{Physical set-up}

\begin{figure}[htb]
\centering
{\includegraphics[width=0.475\textwidth]
{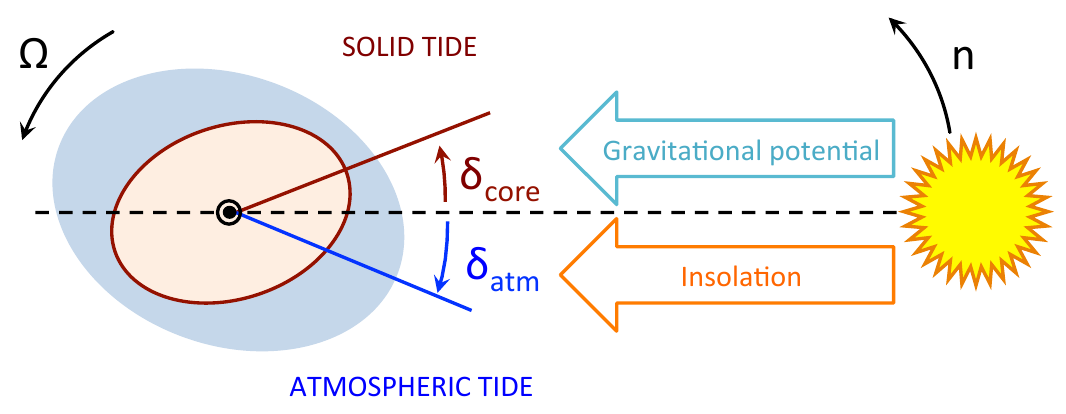}
\textsf{\caption{\label{fig:schema1} Tidal elongation of a Venus-like rotating planet, composed of a solid core (brown) and a gaseous atmosphere (blue), and submitted to gravitational and thermal forcings.} }}
\end{figure}

For simplicity, we consider a planet in a circular orbit of  radius $a$ and mean motion $n$ around a star of mass $ M_* $ and luminosity $ L_* $ exerting on the  planet thermal and gravitational tidal forcing (Fig~\ref{fig:schema1}).
The planet, of mass $ M_{\rm P} $ and spin $ \Omega$, has zero obliquity so that the tidal frequency  is $ \sigma = 2 \omega $, where $ \omega = \Omega - n $. It is composed of a telluric core of radius $ R $ covered by a thin atmospheric layer of mass $ M_{\rm A} = f_{\rm A} M_{\rm P} $ and pressure height scale $ H $ such that $ H \ll R $. This fluid layer is assumed to be homogeneous in composition, of specific \mybf{gas} constant $ \mathcal{R}_{\rm A} =  \mathcal{R}_{\rm GP} / m  $ ($ \mathcal{R}_{\rm GP} $ and $ m $ being the perfect \mybf{gas} constant and the mean molar mass respectively), \mybf{in hydrostatic equilibrium and subject to convective instability, i.e. $ N \approx 0 $, $ N $ designating the Brunt-Väisälä frequency, as observed on the surface of Venus \citep[][]{Seiff1980}}. Hence, the pressure height scale is  
\begin{equation}
H = \frac{\mathcal{R}_{\rm A} T_0}{g},
\end{equation}
where $ T_0 $ is the equilibrium \mybf{surface} temperature of the atmosphere and $ g $ the surface gravity which are related to the equilibrium radial distributions of density $ \rho_0 $ and pressure $ p_0 $ as $ p_0 = \rho_0 g H $. We introduce the first adiabatic exponent of the \mybf{gas} $ \Gamma_1 = \left( \partial \ln p_0 / \partial \ln \rho_0 \right)_{\rm S} $ (the subscript $ S $ being the specific macroscopic entropy) and the parameter \mybf{$ \kappa = 1  - 1 / \Gamma_1 $}. The radiative losses of the atmosphere, \mybf{treated as a Newtonian cooling}, \mybf{lead us} to define a radiative frequency $ \sigma_0 $, given by
\begin{equation}
J_{\rm rad} = C_p \sigma_0 \delta T,
\end{equation}
where \mybf{$ C_p = \mathcal{R}_{\rm A} / \kappa $ is the thermal capacity of the atmosphere per unit mass and}  $ J_{\rm rad} $ is the radiated power per unit mass caused by the temperature variation $ \delta T $ around the equilibrium state. \mybf{It shall be noticed here that the Newtonian cooling describes the radiative losses of optically thin atmospheres, where the radiative transfers between layers can be ignored. We apply this modeling to optically thick atmosphere, such as Venus' one, because the numerical simulations by \cite{Leconte2015} show that it \mybf{can also describe} well tidal dissipation in these cases, with an effective Newtonian cooling frequency.} For more details about this physical set-up, we refer the reader to P1.

\subsection{Atmospheric tidal torque}

For null obliquity and eccentricity, the tidal gravitational potential is reduced to the quadrupolar term of the multipole expansion \citep{Kaula1964}, $ U = U_2 \left( r \right) P_2^2 \left( \cos \theta \right) e^{i \left( \sigma t + 2 \varphi \right) } $, where $ t $ is the time, $ \varphi $ the longitude, $ \theta $ the colatitude, $ r $ the radial coordinate, $ P_2^2 $ the normalized associated Legendre polynomial of order $ \left( l , m \right) = \left( 2 , 2 \right) $ and $ U_2 $ its radial function. Since $ H \ll R $, $ U_2 $ is assumed to be constant. \mybf{The thick atmosphere of \cora{a Venus-like} planet absorbs most of the stellar flux in its upper regions, which are, as a consequence, strongly thermally forced; only 3 \% of the flux reaches the surface \citep[][]{PY1975}. However the tidal effects resulting from the heating by the ground determine the tidal mass redistribution since the atmosphere is far denser near the surface than in upper regions \citep[][]{DI1980,SZ1990}. These layers can also be considered in solid rotation with the solid part as a first approximation because the velocity of horizontal winds is less than $ 5 \ {\rm m.s^{-1}} $ below an altitude of 10 km \citep[][]{Marov1973}. So,} introducing the mean power per unit mass $ J_2 $, we choose for the thermal forcing \mybf{the heating at the ground distribution $ J = J_2 \tau_{\rm J} e^{- \tau_{\rm J} x} P_2^2 \left( \cos \theta \right) e^{i \left( \sigma t + 2 \varphi \right)} $, where $ \tau_{\rm J} \gg 1 $ represents the damping rate of the heating with altitude and depends on the vertical thermal diffusion of the atmosphere at the surface \citep[e.g.][]{CL70}}. This allows us to establish for the atmospheric tidal torque the expression \mybf{(see P1, Eq. 174)}
\begin{equation}
\hat{\mathcal{T}}_{\rm A} = 2 \pi \kappa \frac{R^2 \rho_0 \left( 0 \right)}{g} U_2 J_2 \frac{\sigma}{\sigma^2 + \sigma_0^2},
\label{Torque_a1}
\end{equation}
where $ \rho_0 \left( 0 \right) $ is the density at the ground. \pcor{This function of $ \sigma $ is of the same form as the one given by \cite{ID1978} (Eq.~4).} The quadrupolar tidal gravitational potential and heat power are given by \mybf{(P1, Eq.~74)}
\begin{equation}
\begin{array}{rcl}
   \displaystyle U_2 = \frac{1}{2} \sqrt{\frac{3}{2}} \frac{R^2 \mathcal{G} M_*}{a^3} & \mbox{and} & \displaystyle J_2 = \frac{1}{2}  \sqrt{\frac{3}{2}}  \frac{\alpha \varepsilon R^2 L_*}{M_{\rm A} a^2},
\end{array}
\end{equation}
 where $ \mathcal{G} $ designates the gravitational constant, $ \varepsilon $ the effective fraction of power absorbed by the atmosphere and  $ \alpha $ a shape factor depending on the spatial distribution of tidal heat sources\footnote{
Denoting $ \mathcal{F} \left( \Psi \right) $ the distribution of tidal heat sources as a function of the stellar zenith angle $ \Psi $, the parameter $ \alpha $ is defined as
$ \alpha = \int_0^{\pi} \mathcal{F} ( \Psi ) P_2 ( \cos \Psi ) \sin \Psi d \Psi
$,
where $ P_2(X)=\sqrt{5/8} ( 3 X^2 - 1 )  $ is the normalized Legendre polynomial of order $ n = 2 $.
If we assume a heat source of the form
$\mathcal{F} \left( \Psi \right) =  \cos \Psi$ if  $\Psi \in  [ 0,\frac{\pi}{2} ]$, 
and
$\mathcal{F} \left( \Psi \right) = 0$  else,
we get the shape factor 
$\alpha = 1/8\sqrt{5/2} \approx 0.19$.
}. 
The tidal torque exerted on the atmosphere $\hat{\mathcal{T}}_{\rm A}$ is partly transmitted to the telluric core. The efficiency of this dynamical (viscous) coupling between the two layers is weighted by a parameter $ \beta $ ( $ 0 \leq \beta \leq 1 $). Hence, with $ \omega_0 = \sigma_0 / 2 $, the transmitted torque ${\mathcal{T}}_{\rm A} = \beta\hat{\mathcal{T}}_{\rm A}$ (Eq.~\ref{Torque_a1}) becomes
\begin{equation}
\mathcal{T}_{\rm A} = \frac{3}{32} \frac{\kappa \beta  R^4  \mathcal{G} M_* \alpha \varepsilon L_* }{\mathcal{R}_{\rm A} T_0 \, a^5}   \frac{\omega}{\omega^2 + \omega_0^2}.
\label{tideA}
\end{equation}

\subsection{Solid tidal torque}
For simplicity and because of the large variety of possible rheologies, we assume that the telluric core behaves like a damped oscillator. In this framework, called \mybf{the} ``Maxwell model'', the tidal torque exerted on an homogeneous body can be expressed \citep[e.g.][]{RMZL2012}
%
%
\begin{equation}
\mathcal{T}_{S} =  \frac{3 }{4} \frac{\mathcal{G} M_*^2 R^5}{a^6}  \Im \left\{ k_2 \right\} ,\ \mbox{with} \  \Im \left\{ k_2 \right\} = - \frac{3 K}{2  } \frac{  \sigma_{\rm M} \sigma }{ \sigma_{\rm M}^2 + \left( 1 + K \right)^2 \sigma^2},
\end{equation}
the imaginary part of the second order Love number ($ k_2 $), $ K $ a non-dimensional rheological parameter, $ \sigma_{\rm M} $ the relaxation frequency of the material composing the body, with
\begin{equation}
\begin{array}{rcl}
   \displaystyle  K = \frac{38 \pi}{3} \frac{ R^4 G}{\mathcal{G} M_{\rm P}^2} & \mbox{and} & \displaystyle \sigma_{\rm M} = \frac{G}{\eta},
\end{array}
\end{equation}
 where $ G $ and $ \eta $ are the effective shear modulus and viscosity of the telluric core. Finally, introducing the frequency $ \omega_{\rm M} = \sigma_{\rm M}/ \left( 2 + 2 K \right) $, we obtain
\begin{equation}
\mathcal{T}_{S} = - \frac{9}{8} \frac{\mathcal{G} M_*^2 R^5 K \omega_{\rm M}}{ \left( 1 + K \right) a^6} \frac{\omega}{\omega^2 + \omega_{\rm M}^2}.
\label{tideS}
\end{equation}

\section{Rotational equilibrium states}
\subsection{Theory}
\begin{figure}[]
\centering
{\includegraphics[width=0.475\textwidth]
{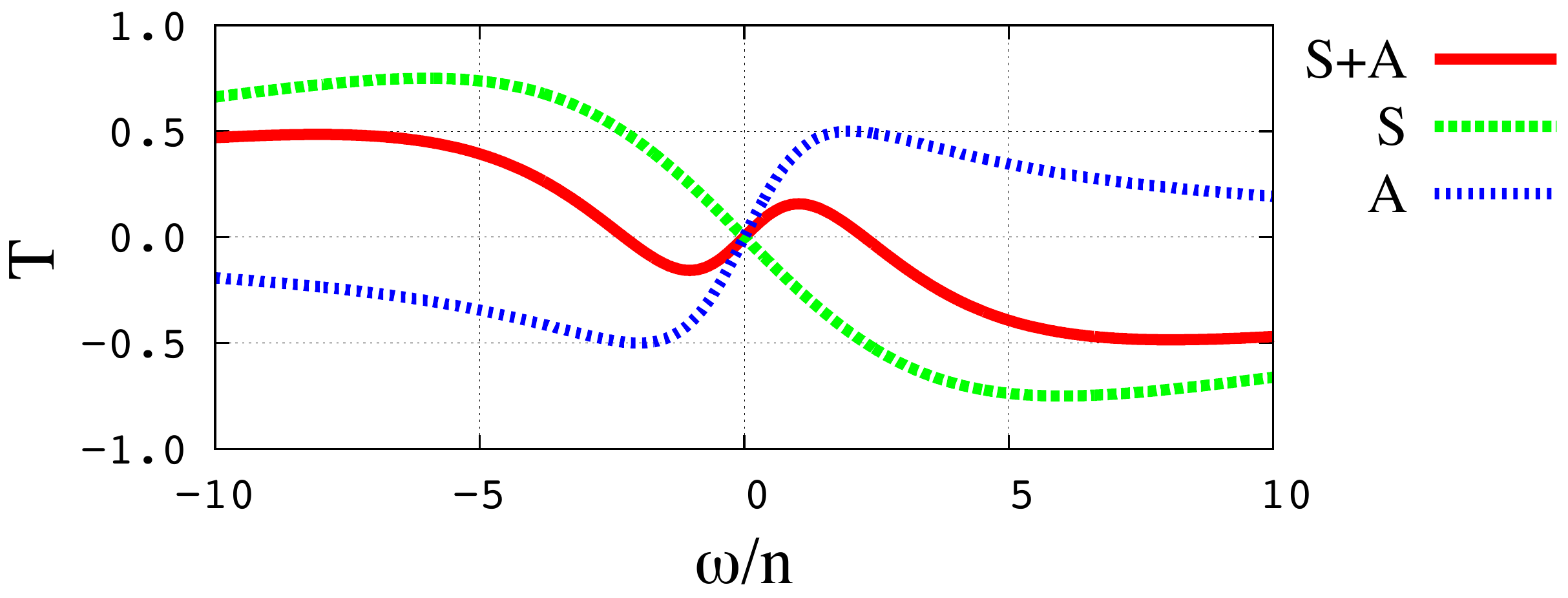}
\textsf{\caption{\label{fig:exemple_couples} Normalized total tidal torque exerted on a Venus-like planet (in red) and its solid (in green) and atmospheric (in blue) components as functions of the forcing frequency $ \omega/n $ with $ \omega_0/n = 2 $ and $ \omega_{\rm M} / n = 6 $.} }}
\end{figure}

The total torque exerted on the planet is the sum of the two previous contributions: $\mathcal{T}_{\rm S+A}  = \mathcal{T}_{\rm S} + \mathcal{T}_{\rm A}$.
When the atmospheric and solid torques are of the same order of magnitude, several \mybf{equilibria} can exist, corresponding to $ \mathcal{T}_{\rm S+A} = 0 $  (Fig.~\ref{fig:exemple_couples}). The synchronization is given by $ \Omega_0 = n $ and non-synchronized retrograde and prograde states of equilibrium, denoted $ \Omega_{-} $ and $ \Omega_{+} $ respectively, are expressed as functions of $ a $ (Eqs.~\ref{tideA},\ref{tideS})
\begin{equation}
\begin{array}{rcl}
    \Omega_{- } \left( a \right) = n - \omega_{\rm eq} \left( a \right)  & \mbox{and} &  \Omega_{+} \left( a \right) =  n + \omega_{\rm eq} \left( a \right) ,
\end{array}
\label{Omega_pm}
\end{equation}
\mybf{where the difference to synchronization, $ \omega_{\rm eq} $, is given by}
\begin{equation}
\begin{array}{rcl}
\displaystyle \omega_{\rm eq} \left( a \right) = \omega_{\rm M} \sqrt{ \frac{a - A \omega_0^2 / \omega_{\rm M}}{A \omega_{\rm M} - a} } & \! \! \! \mbox{with} & \! \! \! \displaystyle A = 12 \frac{ \mathcal{R}_{\rm A} T_0 M_* R K}{\alpha \varepsilon L_* \kappa \beta \left( 1 + K \right)}. 
 \end{array}
 \label{weq}
\end{equation}

\mybf{As a first approximation, we consider that parameters $ A $, $ \omega_0 $ and $ \omega_{\rm M} $ do not vary with the star-planet distance. In this framework,} the equilibrium of synchronization is defined for all orbital \mybf{radii}, \mybf{while} those associated with Eq.~(\ref{weq}) only exist in a zone delimited by $ a_{\rm inf} < a < a_{\rm sup} $, \mybf{the boundaries being}
\begin{equation}
\begin{array}{rcl}
  a_{\rm inf} = A \omega_0  \lambda^{-1}  & \mbox{and} &  a_{\rm sup} = A \omega_0  \lambda 
\end{array}
\label{abd}
\end{equation}
with the ratio $ \lambda = \max \left( \omega_0/\omega_{\rm M} ,\omega_{\rm M}/\omega_0  \right) $. In the particular case where $ \omega_0 = \omega_{\rm M} $, the distance \mybf{$ a_{\rm eq} = A \omega_0  $} corresponds to an orbit for which the atmospheric and solid tidal torques counterbalance each other whatever the angular velocity ($ \Omega $) of the planet. Studying the derivative of $ \omega_{\rm eq} $ (Eq.~\ref{weq}), we note that when the star-planet distance increases,
\begin{itemize}
   \item[$\bullet$] if $ \omega_{\rm M} < \omega_0 $, the pseudo-synchronized states of equilibrium get closer to the synchronization, 
   \item[$\bullet$] if $ \omega_{\rm M} > \omega_0 $, they move away from it.
\end{itemize}

To determine the stability of the identified equilibria, we introduce the first order variation $ \delta \omega $ such that $ \omega = \omega_{\rm eq} + \delta \omega $ and study the sign of $ \mathcal{T}_{\rm S+A} $ at the vicinity of an equilibrium position for a given $ a $. We first treat the synchronization, for which 
\begin{equation}
\begin{array}{rcl}
 \mathcal{T}_{\rm S+A} \,  \propto \, {\rm sign} \left( a - a_{\rm eq} \right) \delta \omega & \mbox{with} & \displaystyle a_{\rm eq} = A \frac{\omega_0^2}{\omega_{\rm M}}.
 \end{array}
\label{stab1}
\end{equation}
Note that $ a_{\rm eq} = a_{\rm inf} $ if $ \omega_0 < \omega_{\rm M} $ and $ a_{\rm eq} = a_{\rm sup} $ \mybf{otherwise}. At the vicinity of non-synchronized equilibria, we have
\begin{equation}
   \mathcal{T}_{\rm S+A} \, \propto \, n^4 \omega_{\rm M} \frac{\omega_{\rm eq}^2 \left( \omega_0^2 - \omega_{\rm M}^2 \right)}{\left(  \omega_{\rm eq}^2 + \omega_{\rm M}^2  \right)^2 \left( \omega_{\rm eq}^2 + \omega_0^2 \right)  } \delta \omega.
\label{stab2}
\end{equation}
Therefore, within the interval $ a \in \left] a_{\rm inf} , a_{\rm sup} \right[ $,
\begin{itemize}
   \item[$\bullet$] if $ \omega_{\rm M} < \omega_0 $, the synchronized state of equilibrium is stable and the pseudo-synchronized ones are unstable,
   \item[$\bullet$] if $ \omega_{\rm M} > \omega_0 $, the synchronized state of equilibrium is unstable and the pseudo-synchronized ones are stable.
\end{itemize}
For $ a < a_{\rm inf} $ the gravitational tide predominates and the equilibrium at $ \Omega = n $ is stable. \mybf{It becomes unstable for $a > a_{\rm sup}$, because the torque is driven by the thermal tide. }
\begin{figure*}[htb]
\centering
{\includegraphics[width=0.30\textwidth]{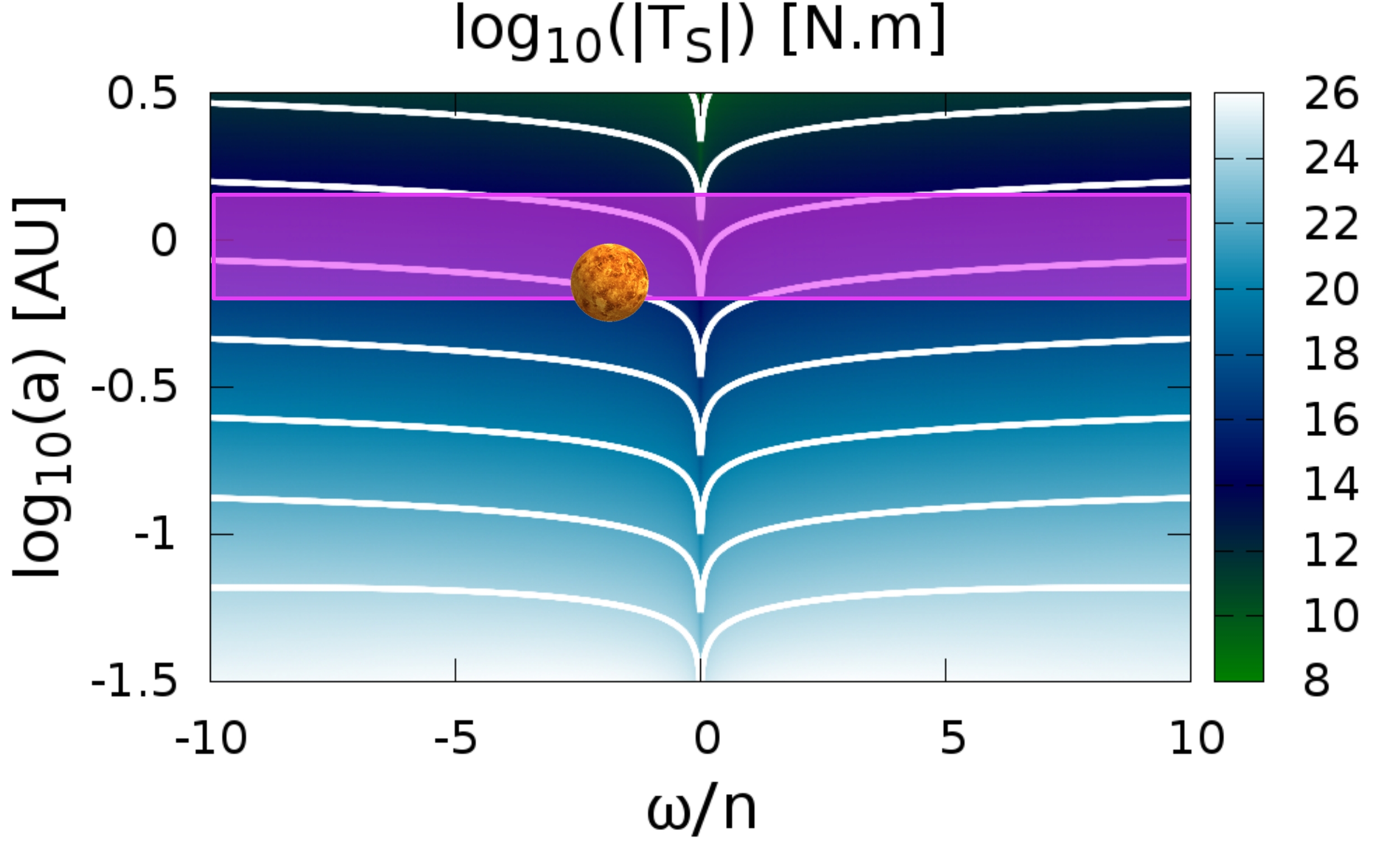} \hspace{0.5cm}
\includegraphics[width=0.30\textwidth]{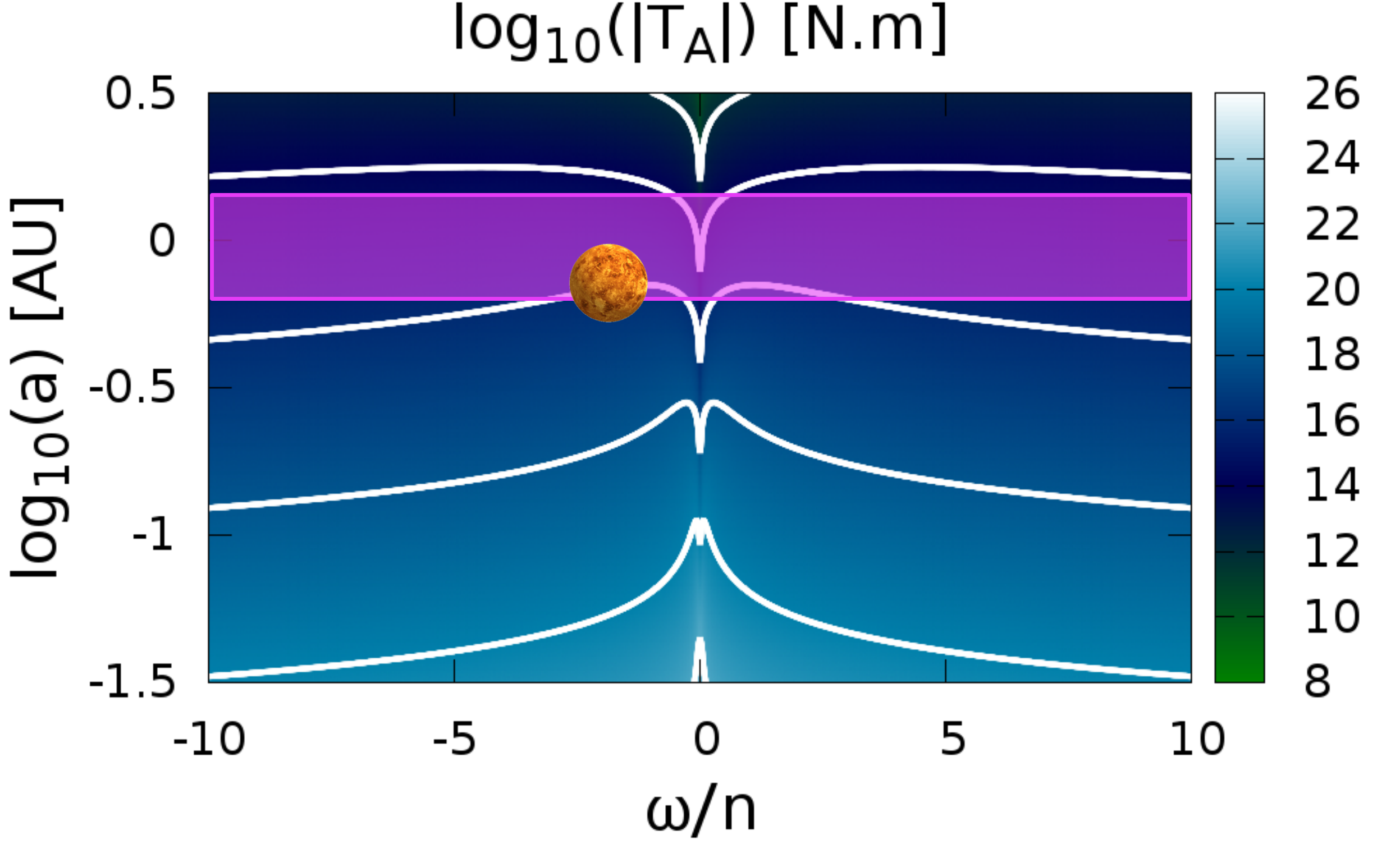}  \hspace{0.5cm}
\includegraphics[width=0.30\textwidth]{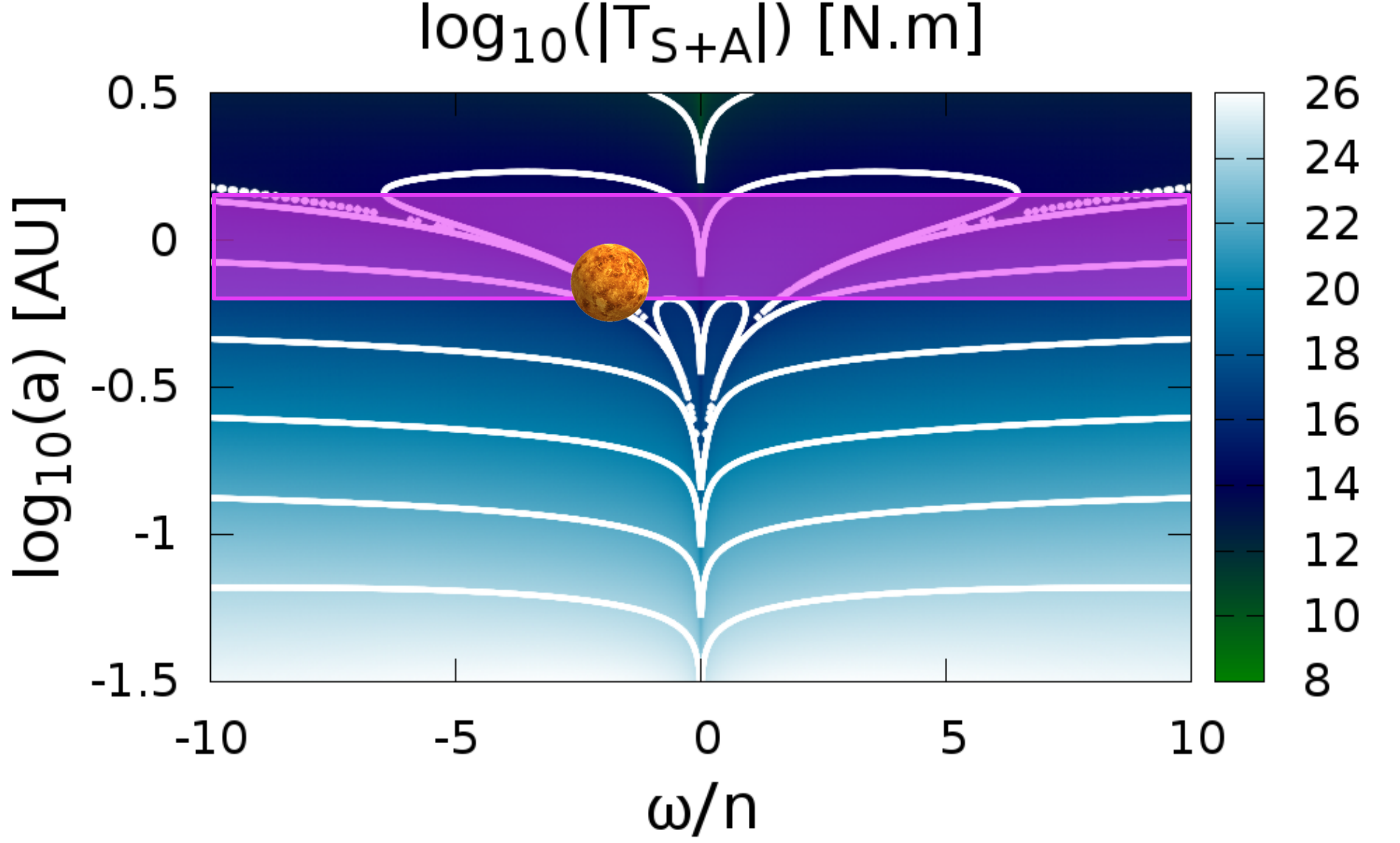} \\[0.5mm]
\includegraphics[width=0.30\textwidth]{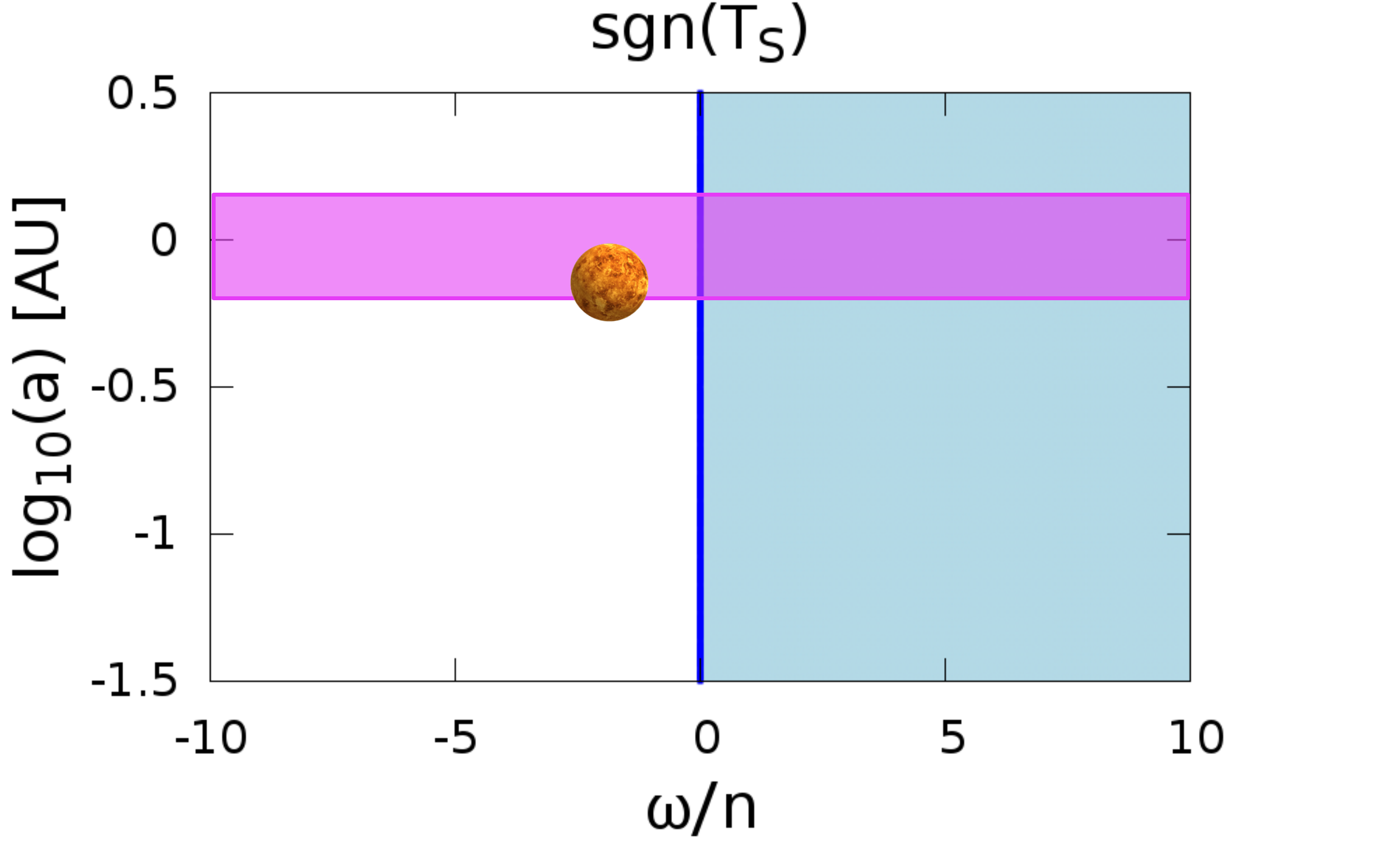}  \hspace{0.5cm}
\includegraphics[width=0.30\textwidth]{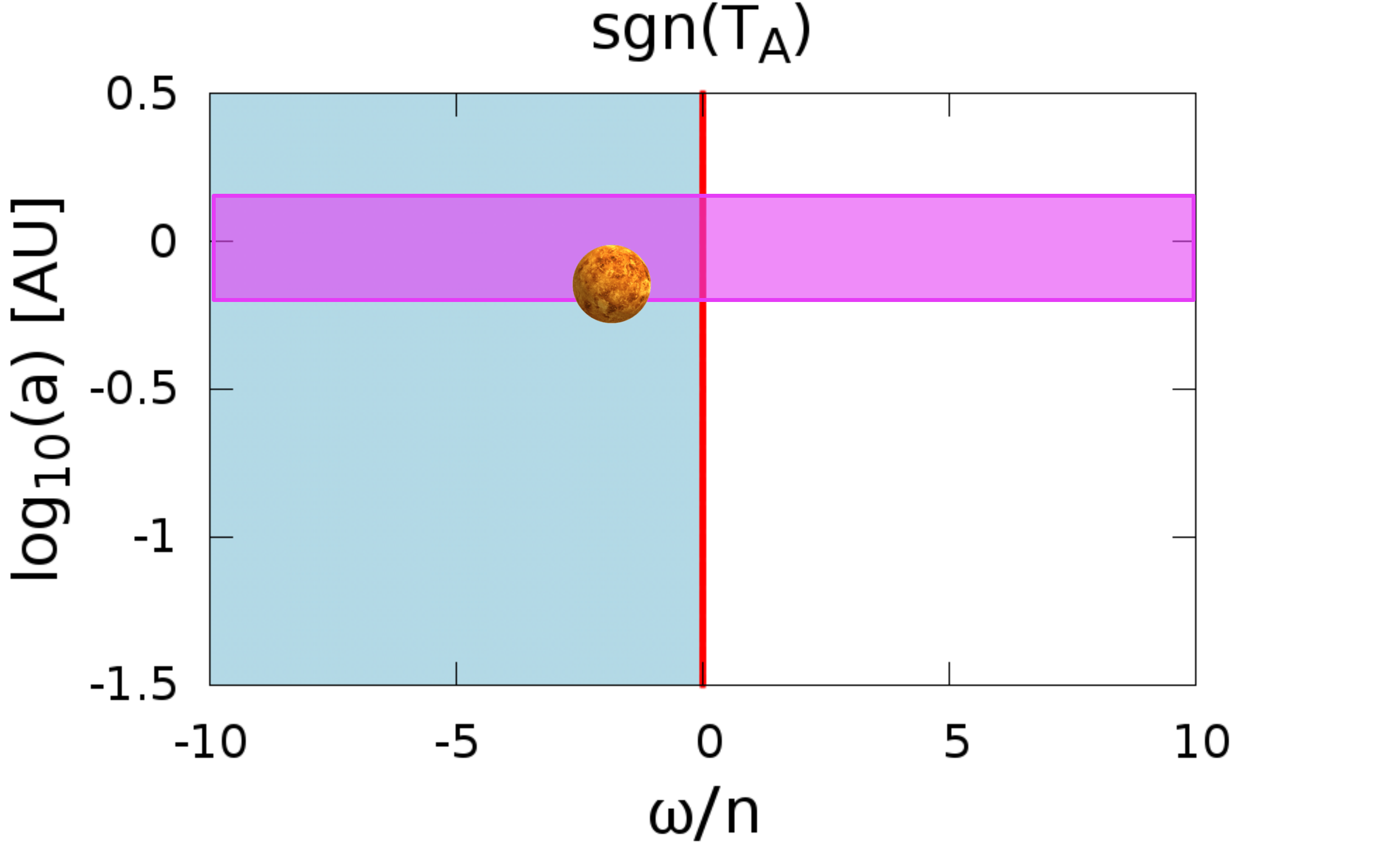}  \hspace{0.5cm}
\includegraphics[width=0.30\textwidth]{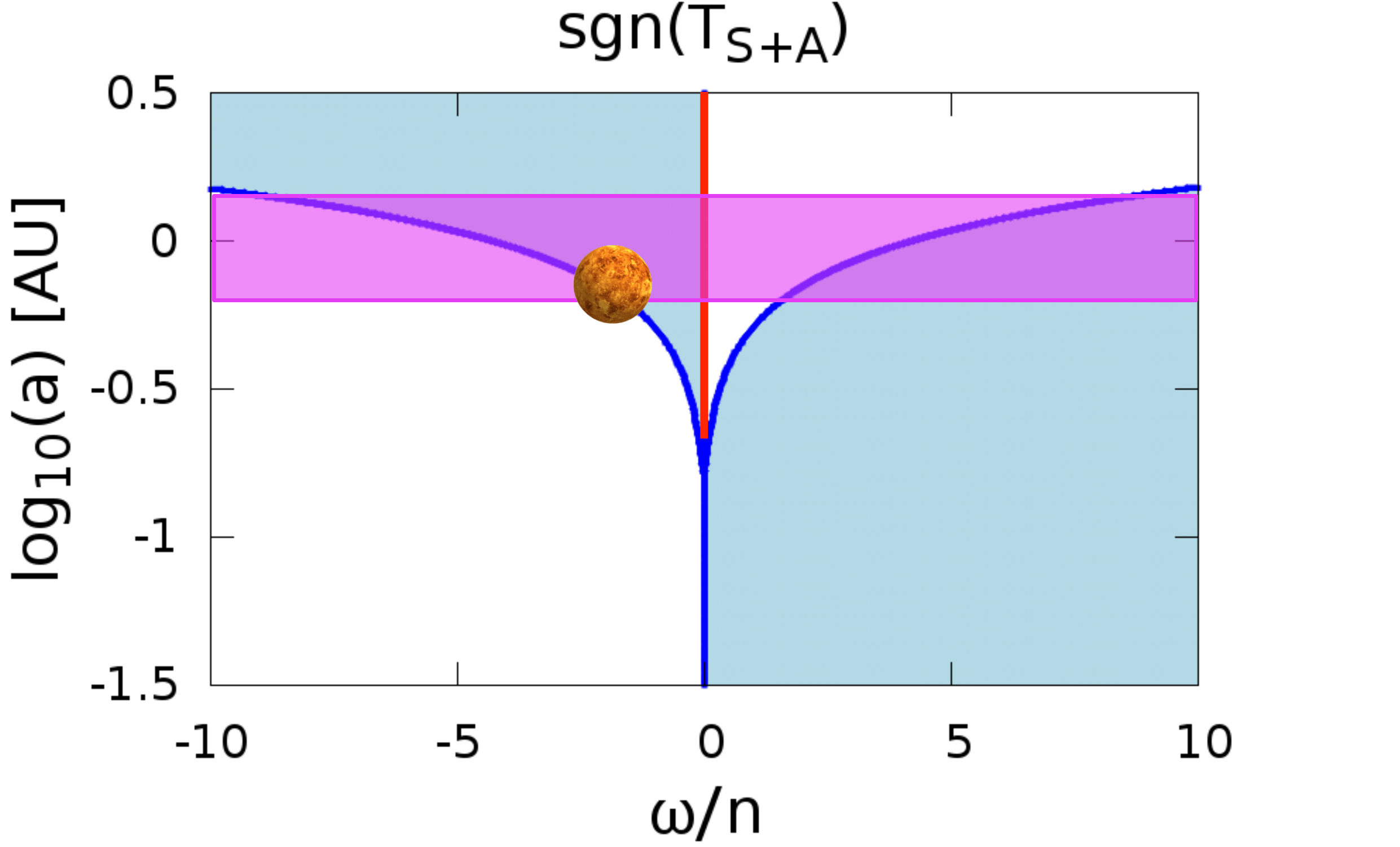} 
\textsf{\caption{\label{fig:Venus} {\it Top:} Solid (left panel), atmospheric (middle panel) and total tidal torque (right panel) exerted on a Venus-like planet as functions of the reduced forcing frequency $ \omega / n $ (horizontal axis) and orbital radius $ a $ in logarithmic scale (vertical axis). The color level corresponds to the torque in logarithmic scale with isolines at $ \log_{10} \left( \mathcal{T} \right) = 8, 10, 12, \ldots, 26 $. {\it Bottom:} Sign of the solid (left panel), atmospheric (middle panel) and total (right panel) tidal torque as functions of the same parameters: white (blue) areas are associated with positive (negative) torques. Stable (unstable) states of equilibrium are \cora{designated} by blue (red) lines \mybf{(with $ A = 1.88 \times 10^{19} \ {\rm m.s} $)}. The pink band corresponds to the habitable zone for the black body equilibrium temperature $ T_{\rm eq} = 288 {\rm K} \pm 20 \% $ for a $ 1 M_\sun $ Solar-type star at the age of the Sun.} }}
\end{figure*}

\subsection{Comparison with previous works}
\label{subsec:comparison_works}
In many early studies dealing with the spin rotation of Venus, the effect of the gravitational component on the atmosphere is ignored so that the atmospheric tide corresponds to a pure thermal tide \citep[e.g.][]{DI1980,CL01,Correia2003,Correia2008,Cunha2015}. Moreover, the torque induced by the tidal elongation of the solid core is generally assumed to be linear in these works. This amounts to considering that $ \left| \omega  \right| \ll \omega_{\rm M} $ and $ \omega_0 \ll \omega_{\rm M} $. Hence, the relation Eq.~(\ref{weq}) giving the positions of non-synchronized equilibria can be simplified:
\begin{equation}
 \displaystyle \omega_{\rm eq} = \sqrt{ \frac{\omega_{\rm M}}{A} a - \omega_0^2 }.
\end{equation}

Following \cite{Correia2008}, we take $ \omega_0 = 0 $. We then obtain $ \omega_{\rm eq} \, \propto \, \sqrt{a} $ while \cite{Correia2008} found $ \omega_{\rm eq} \, \propto \, a $. This difference can be explained by the linear approximation of the \mybf{sine} of the phase lag done in this previous work. Given that the condition $ \omega_0 < \omega_{\rm M} $ is satisfied, the non-synchronized states of equilibrium are stable in this case. 

By using a Global Circulation Model (GCM), \cite{Leconte2015} obtained numerically for the atmospheric tidal torque an expression similar to the one given by Eq.~(\ref{Torque_a1}). They computed with this expression the possible states of equilibrium of Venus-like planets by applying to the telluric core the Andrade and constant-Q models and showed that for both models, \cora{up to} five equilibrium positions could appear. This is not the case when the Maxwell model is used, as shown by the present work. Hence, although these studies agree well on the existence of several non-synchronized states of equilibrium, they also remind us that the theoretical number and positions of \cora{these equilibria} depend on the models used to compute tidal torques and more specifically on the rheology chosen for the solid part. 

%

\begin{table}[h]
\centering
    \begin{tabular}{ l l l }
      \hline
      \hline
      \textsc{Parameters} & \textsc{Values} & \textsc{Units} \\ 
      \hline 
      
      $ \mathcal{G} $ & $ 6.67384 \times 10^{-11}  $ & $ {\rm m^3. kg^{-1}. s^{-2} } $  \\
      $ \mathcal{R}_{\rm GP} $ & $ 8.314 $ & $ {\rm J. mol^{-1} K^{-1} } $  \\
      $ L_* $ & $ 3.846 \times 10^{26} $ & ${\rm W}$  \\ 
      $ \mathcal{G} M_*  $ & $ 1.32712 \times 10^8 $ & ${\rm km^3.s^{-2} }$ \\
      $ R $ & $ 6051.8 $ & $ {\rm km} $  \\
      $ M_{\rm P} $ & $ 4.8676 \times 10^{24} $ & $ {\rm kg} $   \\
      $ f_{\rm A}  $ & $ 10^{-4} $ & --   \\
      $ \mathcal{M}_{\rm A} $ & $ 43.45 $ &  $ {\rm g.mol^{-1}} $  \\
      $ T_0 $ & $ 737 $ & K \\
      $ \kappa \ $ & $ 0.286 $  & -- \\
      $ \omega_0 $ & $ 3.77 \times 10^{-7} $ & $ {\rm s^{-1}} $\\
      $ \alpha $ & $ 0.19 $ & -- \\
      $ \beta $ & $ 1.0 $ & -- \\
      $ \varepsilon $ & $ 0.04 $ & -- \\
      $ G $ & $ 10^{11} $ & Pa \\
      $ \omega_{\rm M}$ & $ 1.075 \times 10^{-4} $  & ${\rm s^{-1}}$ \\
      \hline
    \end{tabular}
    \textsf{\caption{\label{parameters} Numerical values used in the case of Venus-like planets. Parameters from $ \mathcal{G} $ to $ T_0 $ are given by Nasa fact sheets\protect\footnotemark and \cite{codata2010}, the value of $ \kappa $ corresponds to \mybf{a} perfect gas, $ \omega_0 $ is the radiative frequency prescribed by \cite{Leconte2015}  for Venus, $ \alpha $ is computed for a tidal heat power per unit mass proportional to $ \cos \Psi $ on the day side and equal to zero on the night side, the atmospheric tidal torque is assumed to be entirely transmitted to the telluric core ($ \beta = 1 $), $ \varepsilon $ is consistant with the estimation of \cite{DI1980} for the effective heating of Venus' atmosphere by the ground, and \mybf{we take for} $ G $ the shear modulus of silicates \citep[][]{RMZL2012}.   }}
 \end{table}

\section{The case of Venus-like planets}
We now illustrate the  previous results for Venus-like planets (Table~\ref{parameters}). The frequency $ \omega_{\rm M} $ is adjusted so that the present angular velocity of Venus corresponds to the retrograde state of equilibrium identified in the case where the condition $ \omega_{\rm M} > \omega_0  $ is satisfied ($ \Omega_{-} $ in Eq.~\ref{Omega_pm}). In Fig.~\ref{fig:Venus}, we plot the resulting tidal torque and its components, as well as their signs, as functions of the tidal frequency and orbital radius. These maps show that the torque varies over a very large range, particularly the solid component ($ \mathcal{T}_{\rm S} \, \propto \, a^{-6} $ while $ \mathcal{T}_{\rm A} \,\propto \, a^{-5} $). The combination of the solid and atmospheric responses generates the non-synchronized states of equilibrium observed on the bottom left panel, which are located in the interval $ \left] a_{\rm inf} , a_{\rm sup} \right[ $ (see Eq.~\ref{abd}) and move away from the synchronization when $ a $ increases (see Eq.~\ref{weq}), as predicted analytically. 

\begin{figure}[htb]
\centering
{
\includegraphics[width=0.4\textwidth]{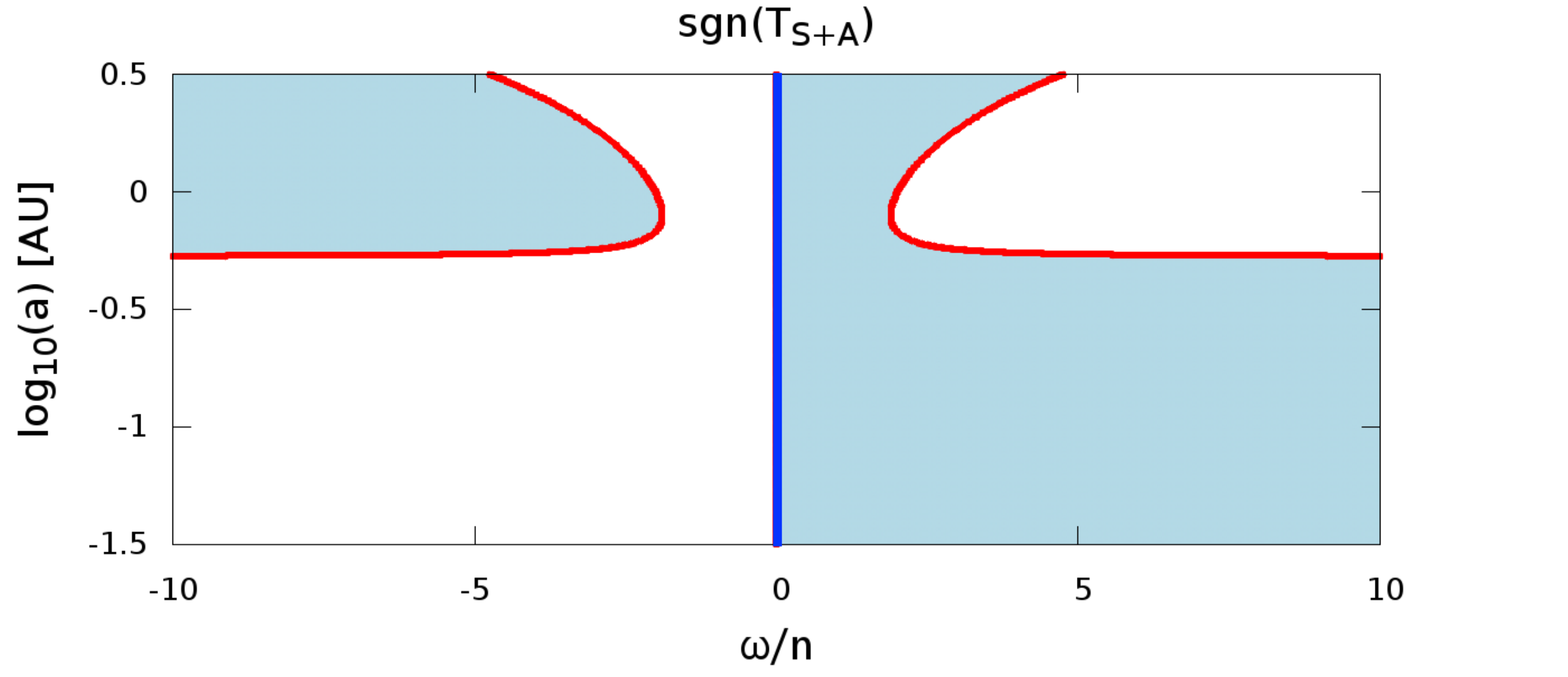} 
\includegraphics[width=0.4\textwidth]{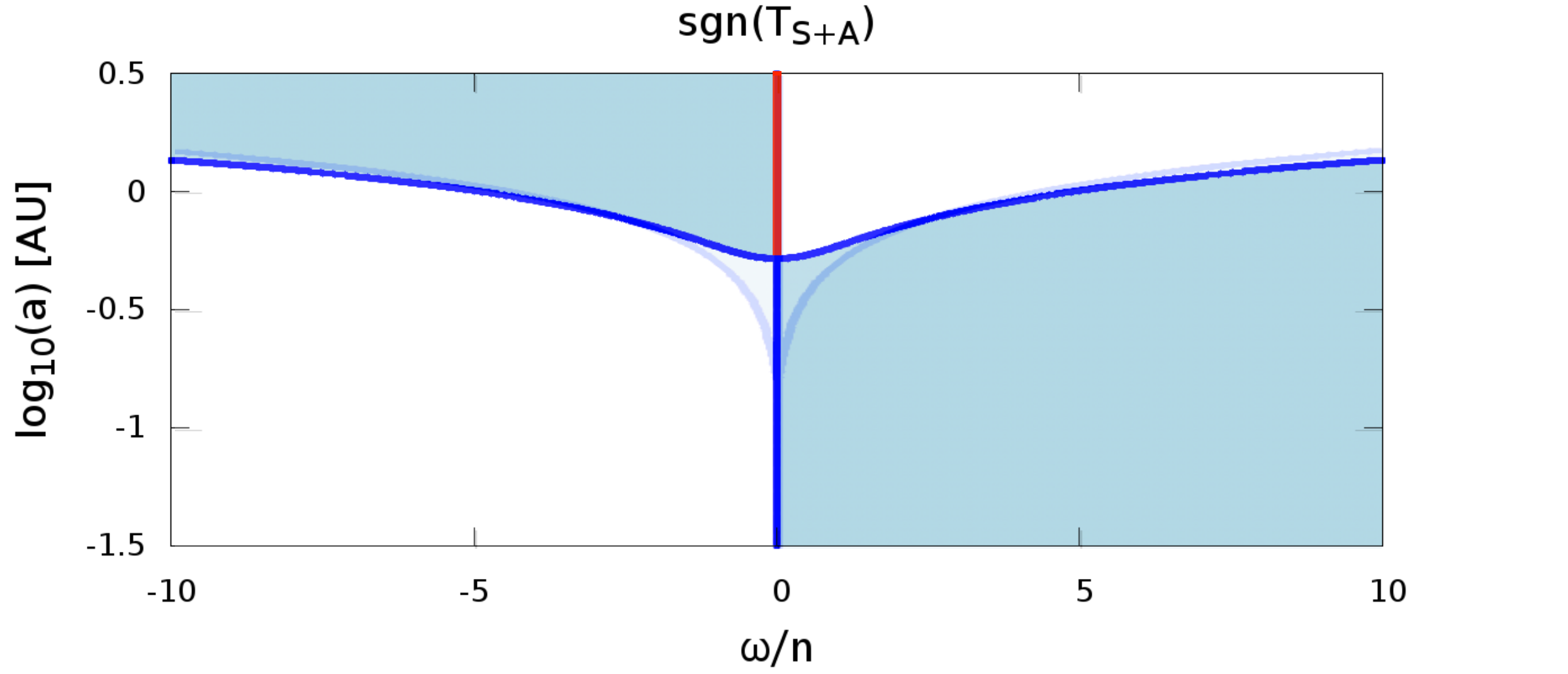} 
\textsf{\caption{\label{fig:explo} \mybf{Sign of the total tidal torque as a function of the tidal frequency $ \omega / n $ (horizontal axis) and orbital radius ($a$)  (vertical axis) for two different cases. {\it Top:}} with $ \omega_0 = 3.77 \times 10^{-7} \ {\rm s^{-1}} $ and 
 $ \omega_{\rm M} =  3.7 \times 10^{-8} \ {\rm s^{-1}} $ ($ \omega_0 > \omega_{\rm M} $). \mybf{{\it Bottom:} with $ T_0 $ and $ \omega_0 $ depending on the star-planet distance \mybf{and $ \omega_{\rm M} > \omega_0 $}. White (blue) areas correspond to positive (negative) torque. Blue (red) lines \cora{designate} stable (unstable) states of equilibrium.}
 }}}

\end{figure}

%
For illustration, \mybf{in Fig.~\ref{fig:explo}}, \cora{we show} the outcome of \cora{a different value} of $ \omega_{\rm M} $, with $ \omega_0 > \omega_{\rm M} $, contrary to 
\mybf{Fig.~\ref{fig:Venus}} where $  \omega_{\rm M} > \omega_0  $.  We observe the behaviour predicted analytically in Sect.~3. In Fig.~\ref{fig:Venus}, the non-synchronized equilibria are stable and move away from the synchronization \cora{when $ a $ increases, but} they are unstable in Fig.~\ref{fig:explo}. 

\mybf{Note that the value of the solid Maxwell frequency obtained for stable non-synchronized states of equilibrium, $ \omega_{\rm M} = 1.075 \times 10^{-4} \ {\rm s^{-1}} $, is far higher than those of typical solid bodies \citep[$\omega_{\rm M} \sim 10^{-10} \ {\rm s^{-1}}$, see][]{Efroimsky2012}. \cora{The Maxwell model, because of its decreasing rate as a function of $\omega$ (i.e. $\propto \, \omega^{-1}$), underestimates the tidal torque for tidal frequencies greater than $\omega_{\rm M}$, which leads to overestimate the Maxwell frequency when equalizing atmospheric and solid torques}. Using the Andrade model for the solid part \cora{could give} more realistic values of $ \omega_{\rm M} $\cora{,} as proved numerically by \cite{Leconte2015}, \cora{because the decreasing rate of the torque is lower in the Andrade model than in the Maxwell model (i.e. $\propto \, \omega^{-\alpha}$ with $\alpha=0.2-0.3$).}  }

\mybf{Finally, we must discuss the assumption we made when we supposed that the parameters of the system did not depend on the star-planet distance. The surface temperature of the planet and the radiative frequency actually vary with $ a $. If we consider that $ T_0 $ is determined by the balance between the heating of the star and the black body radiation of the planet, then $ T_0 \, \propto \, a^{-1/2} $. As $ \sigma_0 \, \propto \, T_0^3 $, we have $ \sigma_0 \, \propto \, a^{-3/2} $. These new dependences modify neither the expressions of $ \omega_{\rm eq} $ (Eq.~\ref{weq}), nor the stability conditions of the states of equilibrium (Eqs.~\ref{stab1} and \ref{stab2}). However, they have repercussions on the boundaries of the region where non-synchronized states exist. This changes are illustrated by Fig.~\ref{fig:explo} (bottom \cora{panel}), which shows the stability diagram of Fig.~\ref{fig:Venus} (bottom left \cora{panel}) computed with the functions $ T_0 \left( a \right) = T_{0;\Venus} \left( a / a_\Venus \right)^{-1/2} $ and $ \sigma_0 \left( a \right) = \sigma_{0;\Venus} \left( a / a_\Venus \right)^{-3/2} $ ($ a_\Venus $, $ T_{0;\Venus} $ and $ \sigma_{0;\Venus} $ being the semi-major axis of Venus and the constant temperature and radiative frequency of Table~\ref{parameters} respectively).   }


\footnotetext{ \url{http://nssdc.gsfc.nasa.gov/planetary/factsheet/}}

\section{Discussion}

A physical model for solid and atmospheric tides \cora{allows us} to determine analytically the possible rotation equilibria of Venus-like planets and their stability. Two regimes exist depending on the hierarchy of the characteristic frequencies associated with dissipative processes (viscous friction in the solid layer and thermal radiation in the atmosphere). These dissipative mechanisms have an impact on the stability of non-synchronized equilibria, which can appear in the habitable zone since they are caused by solid and atmospheric tidal torques of comparable intensities. This study can be used to constrain the equilibrium rotation of observed super-Earths and therefore to infer the possible climates of such planets. Also here the Maxwell rheology was used. This work can be directly applied to alternative rheologies such as the Andrade model \citep[][]{Efroimsky2012}. \mybf{The modeling used here is based on an approach where the atmosphere is assumed to rotate uniformly with the solid part of the planet \citep{ADLM2016}. As general circulation is likely to play an important role in the atmospheric tidal response, we envisage to examine the effect of differential rotation and corresponding zonal winds of the tidal torque in future studies.}


\begin{acknowledgements}
P. Auclair-Desrotour and S. Mathis acknowledge funding by the European Research Council through ERC grant SPIRE 647383. This work was also supported by the ``exoplanètes'' action from Paris Observatory and by the CoRoT/{\it Kepler} and PLATO CNES grant at CEA-Saclay. A. C. M. Correia acknowledges support from CIDMA strategic project UID/MAT/04106/2013.
\end{acknowledgements}

\bibliographystyle{aa}  
\bibliography{ADLM2016} 

\end{document}